\documentclass[prd,preprint,superscriptaddress,showpacs,byrevtex]{revtex4}
\usepackage{epsfig}

\newcommand{\be}{\begin{equation}}
\newcommand{\ee}{\end{equation}}
\newcommand{\bea}{\begin{eqnarray}}
\newcommand{\eea}{\end{eqnarray}}
\def\simgt{\rlap{\lower 3.5 pt \hbox{$\mathchar \sim$}} \raise 1pt
Ê \hbox {$>$}}

\begin{document}

\title{ String Theory at LHC Using Jet Production From String Regge Excitations vs String Balls }

\author{Gouranga C. Nayak}\email{nayak@max2.physics.sunysb.edu}

\affiliation{ 22 West Fourth Street \#1, Lewistown, Pennsylvania 17044, USA}

\date{\today}

\begin{abstract}

If we find extra dimensions in the second run of the LHC in the $pp$ collisions at $\sqrt{s}$ = 14 TeV, then
the string mass scale $M_s$ can be $\sim$ TeV and we should produce QCD jets in
$2 \rightarrow 2$ partonic collisions via string Regge excitations at the LHC. QCD jets can also be produced
from string balls via thermal radiation at Hagedorn temperature. In this paper we study jet production from
string Regge excitations vs string balls in $pp$ collisions at $\sqrt{s}$ = 14 TeV
at LHC and make a comparison with the standard model QCD jets.
We find that high $p_T$ jet production from string Regge excitations can be larger than that from string balls and from
standard model QCD jets. We also find resonances in the jet production cross section in string Regge excitation scenario which
is absent in the other two scenarios. Hence TeV scale high $p_T$ jets can be a good signature to study string Regge excitations in the
$pp$ collisions at $\sqrt{s}$ = 14 TeV at the LHC.

\end{abstract}
\maketitle

\newpage

\section{Introduction}

If extra dimensions \cite{all,all1} are found at
the LHC in the $pp$ collisions at $\sqrt{s}$ = 14 TeV, then the string mass scale $M_s$ could be around
$\sim$ TeV and we should produce: 1) string Regge excitations \cite{dbrane,2string,dstring},
2) string balls \cite{dimo,nayaks,canada,nayakh} and 3) black holes \cite{ppbf,pp,pp1,cham,ball,ball1}
in the $pp$ collisions at $\sqrt{s}$ = 14 TeV at the LHC.
Note that in the first run at LHC in $pp$ collisions at $\sqrt{s}$ = 7 TeV and 8 TeV we have not found
any experimental evidence of beyond standard model physics at LHC \cite{lhc}. LHC has started its second run
with $pp$ collisions at $\sqrt{s}$ = 13 TeV and it will achieve its maximum energy of $pp$ collsions
at $\sqrt{s}$ = 14 TeV in future. Hence all our analysis in this paper will be at the maximum energy
at LHC, {\it i. e.}, we will perform our calculation in this paper for $pp$ collisions at $\sqrt{s}$ =
14 TeV at LHC. LHC in its second run will also collide two lead nuclei at $\sqrt{s}$ = 5.5 TeV per nucleon
which will achieve the total energy $\sim $ 1150 TeV to produce quark-gluon plasma \cite{qgp,qgp1}
where we may expect to observe new physics \cite{newph}.

The string Regge excitations with masses of order $M_s$ can be searched
in $2 \rightarrow 2$ partonic processes in pp collisions at LHC in the weak coupling limit in a model independent
framework \cite{dstring}. In this case a whole tower of infinite string excitations will open up and
the new particles follow the Regge trajectories of vibrating string
\bea
j= j_0 + \alpha' M^2
\label{j}
\eea
with spin $j$ where the Regge slope parameter $\alpha'$ determines the fundamental string
mass scale \cite{dstring}
\bea
\alpha'=\frac{1}{M^2_{\rm string}}.
\eea
These stringy states will lead to new contributions to standard model
scattering processes. This is based on the extensions of standard model where open strings
ends on D-branes, with gauge bosons due to strings attached to stacks of D-branes and
chiral matter due to strings stretching between intersecting D-branes \cite{dbrane}. Using this idea
recently dijet production amplitude is calculated in $2 \rightarrow 2$ partonic collisions.
Since the string Regge resonances
occur around $M_s$ we expect an enhancement of TeV scale jet production at CERN LHC from string Regge
excitations.

String theory is also studied at LHC via string ball production \cite{dimo,cham,nayaks}.
A string ball is a highly excited long string which emits massless
(and massive) particles at Hagedorn temperature with thermal spectrum \cite{amati,canada}.
In string theory the string ball mass $M_{SB}$ is larger than the string mass scale $M_s$
\cite{susskind,allstring}. Typically
\bea
 M_s < M_{SB}< \frac{M_s}{g_s^2}
\label{scale}
\eea
where $g_s$ is the string coupling which can be less than 1 for the string perturbation theory to be valid.
The Hagedorn temperature of a string ball is given by
\bea
T_{SB}=\frac{M_s}{\sqrt{8} \pi}.
\label{tsb}
\eea
Since this temperature is very high at LHC ($\sim$ hundreds of GeV) we also expect an enhancement
of TeV scale jet production at CERN LHC from string balls.

In this paper we study jet production from string balls and from string Regge excitations at LHC
and make a comparison with the standard model QCD jets. We find that high $p_T$
jet production from string Regge excitations can be larger than that from string balls and from
standard model QCD jets. We also find resonances in the jet production cross section in string Regge
excitation scenario which is absent in the other two scenarios. Hence TeV scale high $p_T$ jets can be a good
signature to study string Regge excitations at LHC.

The paper is organized as follows. In section II we describe QCD jet production via string Regge excitations
in $2 \rightarrow 2$ parton fusion processes at LHC. In section III we describe jet production from string balls at LHC.
We present our results and discussions in section IV and conclude in section V.

\section{ Jet Production in $ 2 \rightarrow 2$ Processes via String Regge Excitations }

The
string Regge excitations with masses of order $M_s$ can be searched
in $2 \rightarrow 2$ partonic processes in pp collisions at LHC in the weak coupling limit in a model independent
framework \cite{dstring}. The $ 2 \rightarrow 2$ partonic scattering amplitudes via string Regge excitations can be computed by using
string perturbation theory. At leading order one finds
\bea
&&~|M(gg \rightarrow gg)|^2 = \frac{19}{12} \frac{16 \pi^2 \alpha_s^2}{M_s^4}  \times
[W_{g^*}^{gg \rightarrow gg }[ \frac{{\hat s}^4}{({\hat s}-M_s^2)^2+(\Gamma_{g^*}^{J=0} M_s)^2}
+\frac{({\hat u}^4+{\hat t}^4)}{({\hat s}-M_s^2)^2+(\Gamma_{g^*}^{J=2} M_s)^2}] \nonumber \\
&& +W_{C^*}^{gg \rightarrow gg } [\frac{{\hat s}^4}{({\hat s}-M_s^2)^2+(\Gamma_{C^*}^{J=0} M_s)^2}
+\frac{({\hat u}^4+{\hat t}^4)}{({\hat s}-M_s^2)^2+(\Gamma_{C^*}^{J=2} M_s)^2}]],
\label{m1}
\eea
\bea
&&~|M(gg \rightarrow q \bar q)|^2 = \frac{7}{24} \frac{16 \pi^2 \alpha_s^2}{M_s^4} N_f \times \nonumber \\
&&~[W_{g^*}^{gg \rightarrow q \bar q} \frac{{\hat u}{\hat t}({\hat u}^2+{\hat t}^2)}{({\hat s}-M_s^2)^2+(\Gamma_{g^*}^{J=2} M_s)^2}
+W_{C^*}^{gg \rightarrow q \bar q} \frac{{\hat u}{\hat t}({\hat u}^2+t^2)}{({\hat s}-M_s^2)^2+(\Gamma_{C^*}^{J=2} M_s)^2}],
\label{m2}
\eea
\bea
&&~|M(q {\bar q} \rightarrow gg )|^2 = \frac{56}{27} \frac{16 \pi^2 \alpha_s^2}{M_s^4}  \times \nonumber \\
&&~[W_{g^*}^{q {\bar q} \rightarrow gg } \frac{{\hat u}{\hat t}({\hat u}^2+{\hat t}^2)}{({\hat s}-M_s^2)^2+(\Gamma_{g^*}^{J=2} M_s)^2}
+W_{C^*}^{q {\bar q} \rightarrow gg } \frac{{\hat u}{\hat t}({\hat u}^2+t^2)}{({\hat s}-M_s^2)^2+(\Gamma_{C^*}^{J=2} M_s)^2}]
\label{m3}
\eea
and
\bea
|M(q g \rightarrow q g )|^2 = -\frac{4}{9} \frac{16 \pi^2 \alpha_s^2}{M_s^2}  \times [\frac{{\hat u}{\hat s}^2}{({\hat s}-M_s^2)^2+(\Gamma_{q^*}^{J=1/2} M_s)^2}+\frac{{\hat u}^3}{({\hat s}-M_s^2)^2+(\Gamma_{q^*}^{J=3/2} M_s)^2}].
\label{m4}
\eea
Here ${\hat s}$, ${\hat t}$ and ${\hat u}$ are the Mandelstam variables at partonic level. For massless partons,
${\hat s}+{\hat t} +{\hat u}=0$ in the $2 \rightarrow 2$ partonic scattering processes. In the above expressions
$\alpha_s$ is the QCD coupling constant and
\bea
&&~W_{g^*}^{gg \rightarrow gg}=0.09,~~~~W_{C^*}^{gg \rightarrow gg}=0.91,~~~~W_{g^*}^{gg \rightarrow q \bar q}=W_{g^*}^{q
{\bar q} \rightarrow gg}=0.24,~~~~W_{C^*}^{gg \rightarrow q \bar q}=W_{C^*}^{q {\bar q} \rightarrow gg }=0.76; \nonumber \\
&&~\Gamma_{g^*}^{J=2}=45 (M_s/{\rm TeV}) {\rm GeV},~~~~~~~~~~\Gamma_{g^*}^{J=0}=\Gamma_{C^*}^{J=2}=75 (M_s/{\rm TeV}) {\rm GeV}, \nonumber \\
&&~\Gamma_{C^*}^{J=0}=150 (M_s/{\rm TeV}) {\rm GeV},~~~~~~~~~~\Gamma_{q^*}^{J=1/2}=\Gamma_{q^*}^{J=1/2}=37 (M_s/{\rm TeV}) {\rm GeV}.
\label{rest}
\eea
The $\frac{d\sigma}{dp_T}$ of jet production at LHC is given by
\bea
\frac{d\sigma}{dp_T} = \sum_{a,b}~\frac{p_T}{{8 \pi s}}~\int dy \int dy_2 ~\frac{1}{{\hat s}}~f_a(x_1, Q^2)~ f_b(x_2,Q^2)~\times~
|M(ab \rightarrow cd)|^2
\label{dsdpt}
\eea
where $a,b=q,{\bar q}, g$ and
\bea
x_1=\frac{p_T}{\sqrt{s}} [e^{y}+e^{y_2}],~~~~~~~~~~~~~~x_2=\frac{p_T}{\sqrt{s}} [e^{-y}+e^{-y_2}].
\label{x12}
\eea
We have used CTEQ6M sets \cite{cteq} for the parton distribution function $f(x,Q^2)$ inside the proton
at LHC. We choose the factorization and renormalization scales to be $Q=p_T$ of the jet.

\section{ Jet Production From String Balls at LHC }

For small string coupling $g_s<1$ the Planck length $l_P$ and the quantum
length scale of the string $l_s$ are related by
\bea
l_P^{n+2} \sim g_s^2 l_s^{n+2}
\label{dlength}
\eea
where $n$ is the number of extra dimensions. According to string theory as black hole shrinks
it reaches the correspondence point \cite{susskind,allstring}
\bea
M \le M_{\rm correspondence} \sim \frac{M_s}{g^2_s},~~~~~~~~~~~~~~~~~~M_s \sim \frac{1}{l_s},
\eea
and makes a transition to a configuration dominated by a highly excited long string
known as string ball. The string ball continues to lose mass by evaporation
at the Hagedorn temperature \cite{amati}. Hence the conventional description of evaporation
in terms of black body radiation can be applied. The average radius of the string ball is given by
\bea
R_{SB} \sim \frac{1}{M_s}\sqrt{\frac{M_{SB}}{M_s}}.
\label{rsb}
\eea

Production of a highly excited string from the collision of two
light string states at high $\sqrt{s}$ can be obtained from the Virasoro-Shapiro
four point amplitude by using string perturbation theory. One finds the amplitude
\bea
A(s,t)=\frac{2\pi g_s^2 \Gamma[-1-\alpha's/4]\Gamma[-1-\alpha't/4]\Gamma[-1-\alpha'u/4]}{\Gamma[2+\alpha's/4]
\Gamma[2+\alpha't/4]\Gamma[2+\alpha'u/4]}
\eea
with
\bea
s+t+u =   -16 /\alpha',~~~~~~~~~~~~~~~~~~~~~~~~~~~~\alpha'=l_s^2.
\eea
The production cross section is
\bea
{\hat \sigma} \sim \frac{\pi {\rm Res} A(\alpha's/4 = N, t=0)}{s} = g_s^2 \frac{\pi^2}{8} \alpha'^2 s.
\label{cs1}
\eea
This cross section saturates the unitarity bounds at around $g_s^2 \alpha' s \sim 1$ (or $s \sim \frac{M_s^2}{g_s^2}$).
This implies that the production cross section for string ball grows with $s$ as in eq. (\ref{cs1}) only for
$M_s << \sqrt{s} << M_s/g_s$ while for $\sqrt{s} >> M_s/g_s$ we find $\sigma_{SB} = l_s^2$ which is constant.
Hence the string ball production cross section in a parton-parton collision is given by
\bea
&&~ {\hat \sigma}_{SB}=g_s^2 \frac{\pi^2}{8} \alpha'^2 s \sim \frac{g_s^2 M_{SB}^2}{M_s^4},~~~~~~~~~~~M_s<<M_{SB}<<M_s/g_s, \nonumber \\
&&~ {\hat \sigma}_{SB} =l_s^2 \sim \frac{1}{M_s^2},~~~~~~~~~~~M_s/g_s<<M_{SB}<<M_s/g^2_s.
\label{sbc}
\eea

If string balls are formed at the LHC then they will quickly evaporate by emitting massless
(and massive) particles at Hagedorn temperature with thermal spectrum \cite{amati,canada,nayaks}.
The emission rate for jets with momentum $\vec{p}$ and energy $E$ from a string ball of temperature
$T_{SB}$ is given by
\be
\frac{dN_{\rm jet}}{d^3p dt }=
\frac{A_n c_n}{32\pi^3}\frac{1}{(e^{\frac{E}{T_{SB}}} \pm 1)}\,,
\label{thermal}
\ee
where $A_n$ is the $d(=n+3)$ dimensional area factor \cite{bv,nayaks,nayakh} and $c_n$ is the multiplicity factor.
$c_n$ = 16 for gluon and $c_n$ = 6 for a single flavor quark. The $+$ ($-$) sign is for quark (gluon) jets.

\begin{figure}[htb]
\vspace{2pt}
\centering{{\epsfig{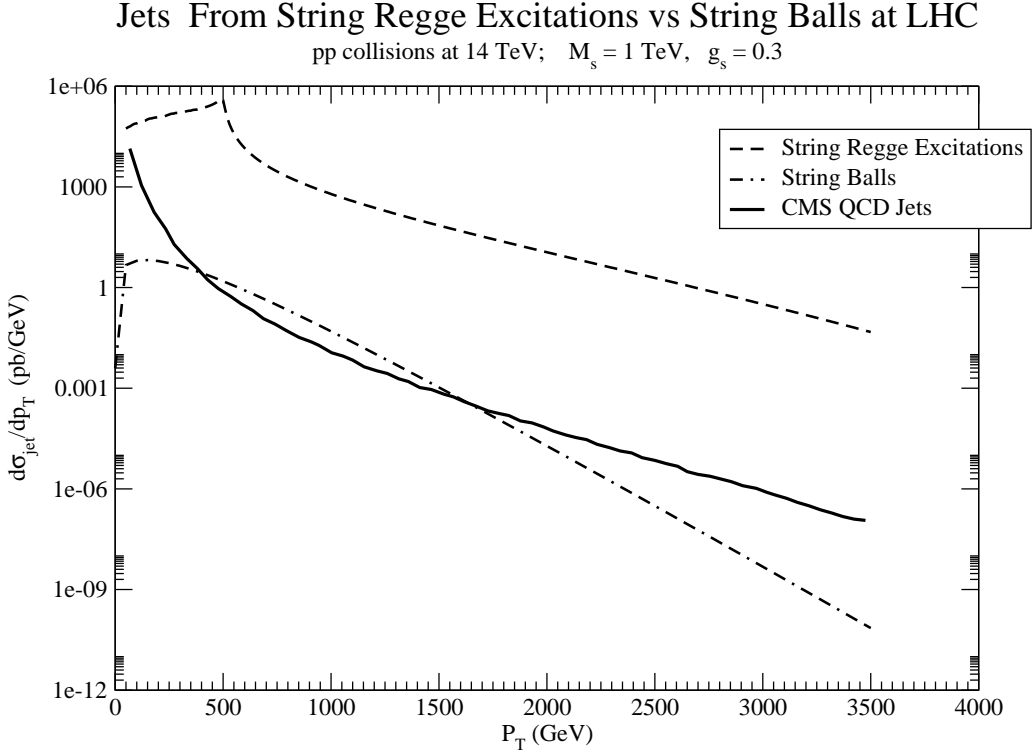}}}
\caption{ $p_T$ distribution of jet production cross section from string Regge excitations, from string balls and from
standard model QCD processes in pp collisions at $\sqrt s$ = 14 TeV at LHC. The string mass scale $M_s$ = 1 TeV.
}
\label{fig1}
\end{figure}

The differential cross section for gluon jet production with momentum $\vec{p}$ and energy
$E =\sqrt{{\vec p}^2}$ from string ball of temperature $T_{SB}$ at LHC is given by \cite{nayaks,arif}
\bea
&& \frac{Ed\sigma_{\rm gluon}}{d^3p}
= \frac{1}{(2\pi)^3s}{\sum}_{ab}~\int_{M^2_{s}}^{\frac{M^2_s}{g^4_s}}~dM^2~
\int \frac{dx_a}{x_a} ~f_{a/p}(x_a, \mu^2)~f_{b/p}(\frac{M^2}{sx_a}, \mu^2)~\hat{\sigma}^{ab}(M)~\frac{A_n c_n \gamma \tau_{SB} p^\mu u_\mu}{(e^{\frac{p^\mu u_\mu}{T_{SB}}} - 1)}\,, \nonumber \\
\label{bksh}
\eea
where $M_s$ is the string mass scale and $g_s$ is the string coupling which is less than $1$ for the string perturbation theory to be valid,
see eq. (\ref{scale}). The flow velocity is $u^\mu$ and $\gamma$ is the Lorentz boost factor with
\bea
\gamma {\vec v}_{SB} = (0,0,\frac{(x_1-x_2) \sqrt{s}}{2M_{SB}}).
\label{4v}
\eea
$A_n$ is the $d(=n+3)$ dimensional area factor \cite{bv,nayaks}. We will use the number of extra dimensions
$n=6$ in our calculation. The partonic level string ball production cross section is given by \cite{dimo}
\bea
&& {\hat \sigma}(M_{SB}) = \frac{1}{M^2_s},~~~~~~~~~~~~~~~~~~~~~~~~\frac{M_s}{g_s}<M_{SB}<\frac{M_s}{g^2_s} \nonumber \\
&& {\hat \sigma}(M_{SB}) = \frac{g^2_sM^2_{SB}}{M^4_s},~~~~~~~~~~~~~~~~~~~~~~~~M_s<M_{SB}<\frac{M_s}{g_s}.
\label{ssb}
\eea
We have used CTEQ6M PDF \cite{cteq} in our calculation.

Similarly, the differential cross section for quark jet production with momentum $\vec{p}$ and energy
$E =\sqrt{{\vec p}^2}$ from string ball of temperature $T_{SB}$ at LHC is given by
\bea
&& \frac{Ed\sigma_{\rm quark}}{d^3p}
= \frac{1}{(2\pi)^3s}{\sum}_{ab}~\int_{M^2_{s}}^{\frac{M^2_s}{g^4_s}}~dM^2~
\int \frac{dx_a}{x_a} ~f_{a/p}(x_a, \mu^2)~f_{b/p}(\frac{M^2}{sx_a}, \mu^2)~\hat{\sigma}^{ab}(M)~\frac{A_n c_n \gamma \tau_{SB} p^\mu u_\mu}{(e^{\frac{p^\mu u_\mu}{T_{SB}}} + 1)}\,. \nonumber \\
\label{gbksh}
\eea

\section{Results and Discussions}

In this section we present results of jet production cross section
in pp collisions at $\sqrt{s}$=14 TeV at LHC from string Regge excitations
and from string balls. We make a comparison with the standard model QCD jets
at the LHC. We use
\bea
g_s=0.3
\eea
in our calculation \cite{canada}.

\begin{figure}[htb]
\vspace{2pt}
\centering{{\epsfig{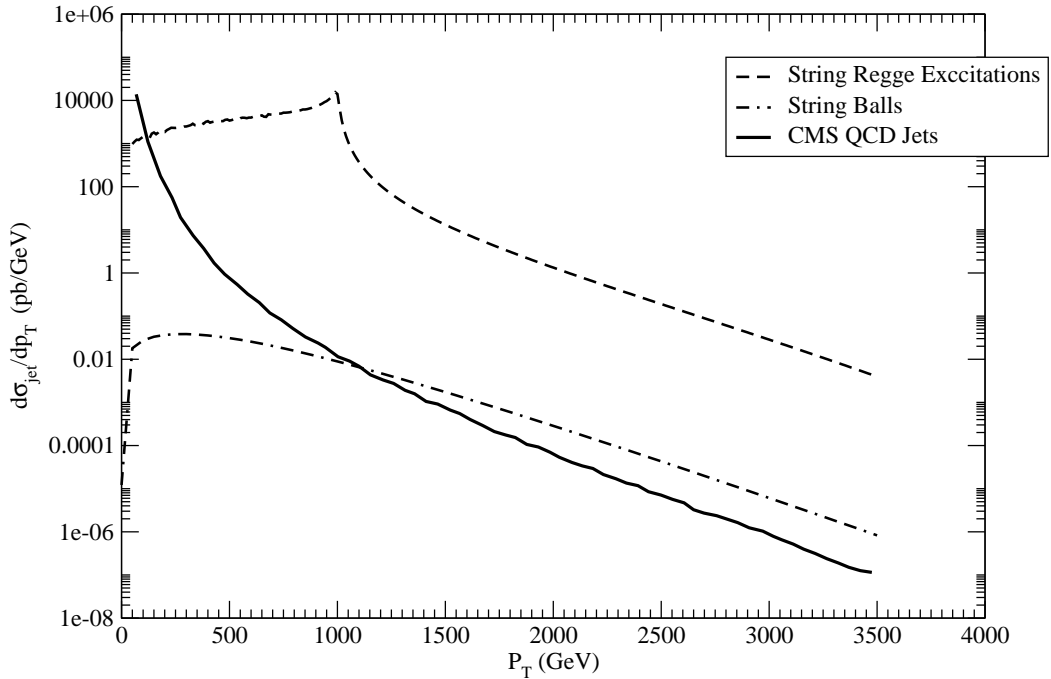}}}
\caption{ $p_T$ distribution of jet production cross section from string Regge excitations, from string balls and from
standard model QCD processes in pp collisions at $\sqrt s$ = 14 TeV at LHC.  The string mass scale $M_s$ = 2 TeV.
}
\label{fig2}
\end{figure}

In Fig. 1 we present $\frac{d\sigma}{dp_T}$ of jet production from string Regge excitations, from string balls and
make a comparison with the standard model QCD jets at the LHC in pp collisions at $\sqrt{s}$=14 TeV.
We use the string mass scale $M_s$ = 1 TeV in our calculation. The dashed line is for jet production from string
Regge excitations at LHC. The dot-dashed line is for jet production from string balls at LHC.
For comparison we present $\frac{d\sigma}{dp_T}$ of standard model QCD jets in the solid line
at LHC from \cite{djet}. It can be seen that if the string mass scale $M_s\sim$ 1 TeV, then
jet production from string Regge excitations is much larger than standard model QCD jets and also that from string balls at LHC.
It can also be seen that there are resonances in the jet production cross section in the string Regge excitation scenario
which is absent in other two scenarios. Hence TeV scale high $p_T$ jet can be a good signature to study string
string Regge excitations at LHC.

\begin{figure}[htb]
\vspace{2pt}
\centering{{\epsfig{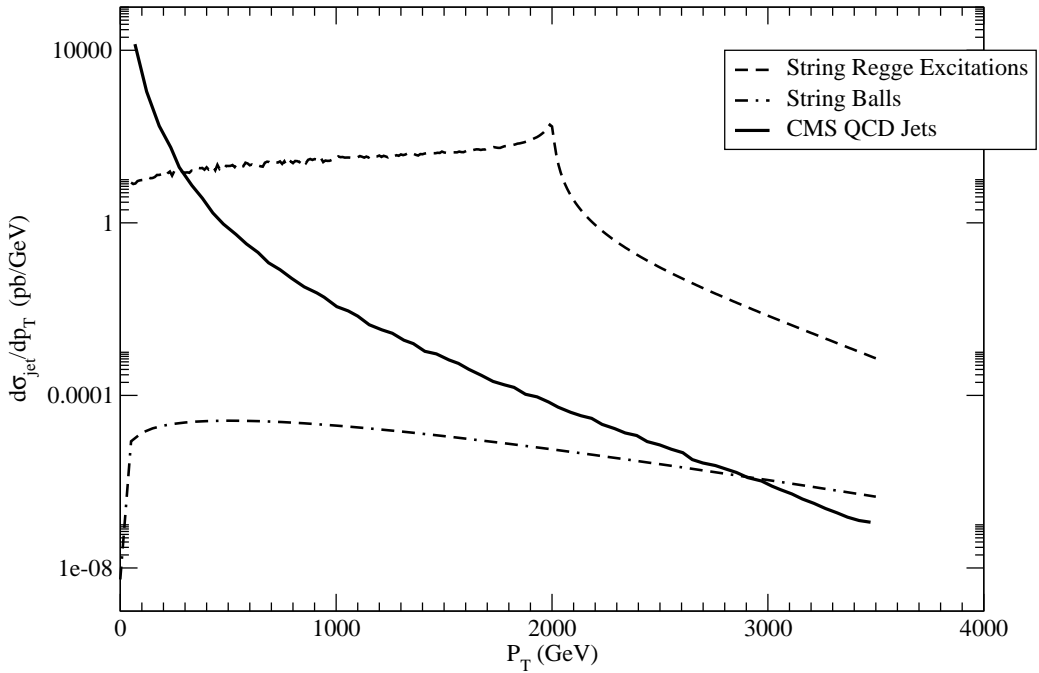}}}
\caption{ $p_T$ distribution of jet production cross section from string Regge excitations, from string balls and from
standard model QCD processes in pp collisions at $\sqrt s$ = 14 TeV at LHC.  The string mass scale $M_s$ = 4 TeV.
}
\label{fig3}
\end{figure}

In Fig. 2 we present $\frac{d\sigma}{dp_T}$ of jet production from string Regge excitations, from string balls and
make a comparison with the standard model QCD jets at the LHC in pp collisions at $\sqrt{s}$=14 TeV
for string mass scale $M_s$ = 2 TeV. The dashed line is for jet production from string
Regge excitations at LHC. The dot-dashed line is for jet production from string balls at LHC.
For comparison we present $\frac{d\sigma}{dp_T}$ of standard model QCD jets in the solid line
at LHC from \cite{djet}. It can be seen that if the string mass scale $M_s\sim$ 2 TeV, then
high $p_T$ jet production ($p_T>$ 100 GeV) from string Regge excitations can be much larger than standard
model QCD jets. It can also be seen that if the string mass scale $M_s\sim$ 2 TeV, then
jet production from string Regge excitations can be much larger than the jet production from string balls
at LHC. For $p_T\lesssim $1 TeV, the standard model jet production is higher than the jet produced from string
balls at LHC, whereas for $p_T\gtrsim $1 TeV the jet production from string balls at LHC is higher than the
jets produced from standard model processes at LHC.
It can also be seen that there are resonances in the jet production cross section in the string Regge excitation scenario
which is absent in other two scenarios. Hence TeV scale high $p_T$ jet can be a good signature to study
string Regge excitations at LHC.

\begin{figure}[htb]
\vspace{2pt}
\centering{{\epsfig{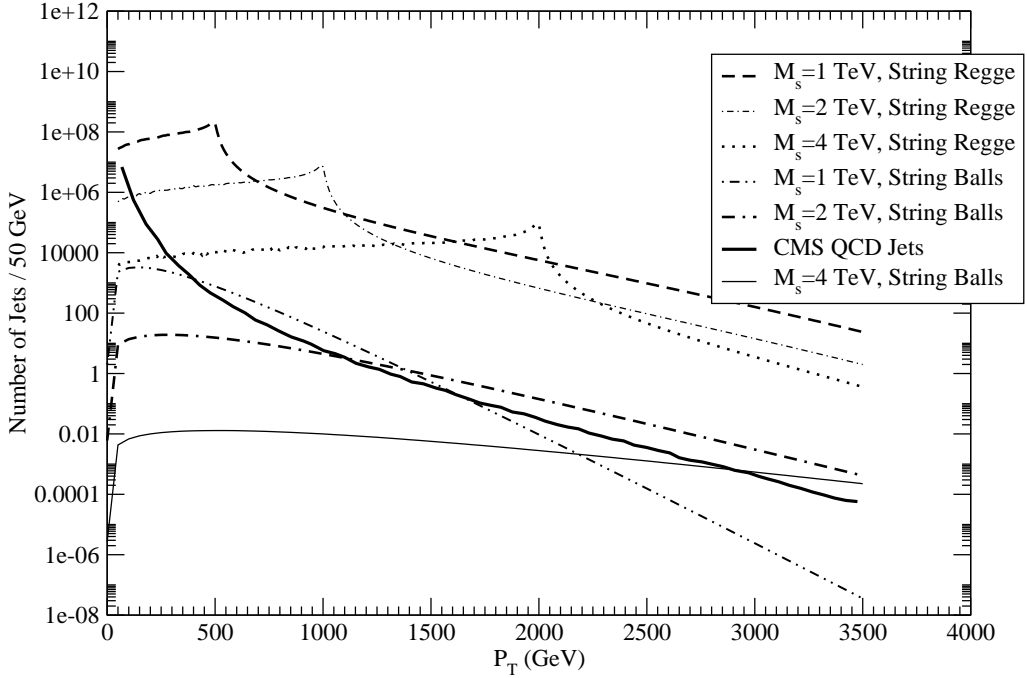}}}
\caption{ $p_T$ distribution of number of jet production from string Regge excitations, from string balls and from
standard model QCD processes in pp collisions at $\sqrt s$ = 14 TeV at LHC with luminosity equals to 10 $pb^{-1}$.
}
\label{fig4}
\end{figure}

In Fig. 3 we present $\frac{d\sigma}{dp_T}$ of jet production from string Regge excitations, from string balls and
make a comparison with the standard model QCD jets at the LHC in pp collisions at $\sqrt{s}$=14 TeV
for string mass scale $M_s$ = 4 TeV. The dashed line is for jet production from string
Regge excitations at LHC. The dot-dashed line is for jet production from string balls at LHC.
For comparison we present $\frac{d\sigma}{dp_T}$ of standard model QCD jets in the solid line
at LHC from \cite{djet}. It can be seen that if the string mass scale $M_s\sim$ 4 TeV, then
high $p_T$ jet production ($p_T>$ 300 GeV) from string Regge excitations can be much larger than standard
model QCD jets. It can also be seen that if the string mass scale $M_s\sim$ 4 TeV, then
jet production from string Regge excitations can be much larger than the jet production from string balls
at LHC. For $p_T\lesssim $3 TeV, the standard model jet production is higher than the jet produced from string
balls at LHC, whereas for $p_T\gtrsim $ 3 TeV the jet production from string balls at LHC is higher than the
jets produced from standard model processes at LHC.
It can also be seen that there are resonances in the jet production cross section in the string Regge excitation scenario
which is absent in other two scenarios. Hence TeV scale high $p_T$ jet can be a good signature to study
string Regge excitations at LHC.

In Fig. 4 we present number of jet production per $p_T$ (in 50 GeV $p_T$ bin) from string Regge excitations,
from string balls for various values of $M_s$ and make a comparison with the standard model QCD jets at LHC.
The dashed, dot-dashed and dotted lines are for jets from string Regge excitations at LHC with string mass scale
$M_s$= 1, 2 and 4 TeV respectively. The dot-dot-dashed, dot-dashed-dashed and thin solid lines are for jets from
string balls at LHC with string mass scale $M_s$ = 1, 2 and 4 TeV respectively.
For comparison we present standard model results for QCD jets in the solid line at LHC from
\cite{djet}. It can be seen that, depending on the values of the string mass scale $M_s$, the jets from string
Regge excitations can be much larger than standard model QCD jets and also that from string balls at LHC. It
can also be seen that there are resonances in the jet production cross section in the string Regge excitation
scenario which is absent in other two scenarios.

Hence we find that TeV scale high $p_T$ jet can be a good signature to study string
string Regge excitations at LHC.

\section{Conclusions}

If we find extra dimensions in the second run of the LHC in the $pp$ collisions at $\sqrt{s}$ = 14 TeV, then
the string mass scale $M_s$ can be $\sim$ TeV and we should produce QCD jets in
$2 \rightarrow 2$ partonic collisions via string Regge excitations at the LHC. QCD jets can also be produced
from string balls via thermal radiation at Hagedorn temperature. In this paper we have studied jet production from
string Regge excitations vs string balls in $pp$ collisions at $\sqrt{s}$ = 14 TeV
at LHC and have made a comparison with the standard model QCD jets.
We have found that high $p_T$ jet production from string Regge excitations can be larger than that from string balls and from
standard model QCD jets. We have also found resonances in the jet production cross section in string Regge excitation scenario which
is absent in the other two scenarios. Hence TeV scale high $p_T$ jets can be a good signature to study string Regge excitations in the
$pp$ collisions at $\sqrt{s}$ = 14 TeV at the LHC.

LHC in its second run will also collide two lead nuclei at $\sqrt{s}$ = 5.5 TeV per nucleon
which will achieve the total energy $\sim $ 1150 TeV to produce quark-gluon plasma \cite{qgp,qgp1,qgp2,qgp3}
where we may expect to observe new physics \cite{newph,amati,bv}.

\end{document}